\documentclass[12pt,a4paper]{article}

 \usepackage[dvips]{graphics}
 \usepackage{graphicx}
 \usepackage{afterpage}
 \usepackage{enumerate}
\begin{document}

 \title{\bf Evolution of solar magnetic tubes \\ and
  its manifestation in Stokes parameters
  }

 \author{\bf V.A. Sheminova and A.S. Gadun}
 \date{}

 \maketitle
 \thanks{}
\begin{center}
{Main Astronomical Observatory, National Academy of Sciences of
Ukraine
\\ Zabolotnoho 27, 03689 Kyiv, Ukraine\\ E-mail: shem@mao.kiev.ua}
\end{center}

 \begin{abstract}
Basic scenarios and mechanisms for the formation and decay of small-scale magnetic
elements and their manifestation in synthesized Stokes profiles of the Fe I 15648.5
\AA\ infrared line are considered in the context of two-dimensional modeling of
nonstationary magnetogranulation on the Sun. The stage of convective collapse is
characterized by large redshifts in the V profiles accompanied by complete Zeeman
splitting of the I profiles. This is due to intense downward flows of material,
which facilitates the concentration of longitudinal field with an amplitude of
about several kG in the tube. The dissipation of strong magnetic structures is
characterized by blueshifts in the Stokes profiles, which result from upward fluxes
that decrease the magnetic field in the tube. Typical signatures during key stages
in the evolution of compact magnetic elements should be detectable via observations
with sufficiently high spatial and temporal resolution.
\end{abstract}

\section{Introduction}
     \label{S-Introduction}

Studies of the interaction of thermal convection with magnetic fields via numerical
modeling of magneto-convection \cite{Atroshchenko, Brandt,  Gadun00,  Gross98,
 Nordlung86, Nordlung90, Steiner}  are of considerable interest for our
understanding of the structure and dynamics of small-scale magnetic fields, which
carry most of the magnetic flux going from the solar photosphere, outside of pores
and sunspots. Current time-dependent magnetohydro-dynamic (MHD) models can
successfully describe the observed concentration of magnetic flux in intergranular
regions and explain variations in convection under the influence of magnetic
fields. They also enable detailed study of mechanisms for the formation of compact
magnetic elements and their evolution. Unfortunately, verification of MHD models
and direct comparison with observations are not easy for a number of reasons, such
as the limited spatial resolution of the observations.  One link between
observations and MHD models is provided by Stokes diagnostics, which are now well
developed. These represent a set of special methods for extracting information
about the structure and dynamics of magnetic elements from spectropolarimetric
observations. The use of Stokes diagnostics to study the structure of small-scale
magnetic elements is necessitated by the fact that these elements are smaller
(70--300~km) than the spatial resolution of modem instruments, so that the observed
signal carries information not only about the magnetic structures, but also about
the nonmagnetic surrounding environment. As a result, correct interpretation of
such observations is difficult.

The aim of the present work is to study evolutionary changes in the structure and
dynamics of magnetic elements, as well as variations in the parameters of
synthesized Stokes profiles, via two-dimensional modeling of nonstationary
magnetogranulation. We also wish to identify diagnostic signatures characteristic
of magnetic-field intensification and dissipation.

The numerical modeling of nonstationary magneto-convection we consider here was
carried out for two hours of real solar time, and describes both the formation and
dissipation of small-scale magnetic elements. In particular, precisely such
modeling demonstrated for the first time the importance of the surface mechanism
(scenario) for the formation of compact magnetic elements, which is realized during
the fragmentation of large-scale thermal fluxes. This mechanism is described in
detail by Gadun et al. \cite{Gadun99}, who also present the preliminary results of
the modeling. These results are analyzed in more detail in \cite{Gadun00}, along
with investigations of the formation and decay of tubes, tube stability regimes,
differences in the brightness characteristics of magnetic and nonmagnetic
granulation, and the dependences of magnetic-element parameters on their horizontal
scales and the field intensity. Extensive test-diagnostic calculations---in
particular, the synthesis of Stokes profiles of several spectral lines and
comparison with observations---were conducted for the same models in
\cite{Sheminova99}. These models adequately describe the basic features of
spectropolarimetric observations of compact magnetic elements. Another important
result of \cite{Sheminova99} was testing the available Stokes diagnostics methods,
which was done by comparing the model parameters with those derived from an
analysis of theoretical spectropolarimetric scans.

The present paper represents a continuation of these studies. It describes in more
detail several critical stages in the evolution of compact magnetic elements that
can be detected via spectropolarimetric methods. As usual, the small-scale magnetic
elements in the numerical modeling will be called tubes, although they do not have
a tube shape in a two-dimensional planar representation.

\section{MHD models and calculation of Stokes profiles
}

We estimated all the thermodynamic parameters of the atmosphere required to solve
the transfer equations for the spectral lines in the presence of a magnetic field
via numerical MHD modeling of planar magnetoconvection of the granulation scales.
We made the approximation that the medium was compressible and partially ionized,
and was stratified by gravity and coupled with the radiation. The complete system
of radiative MHD equations and its solution for the particular case at hand are
described in \cite{Gadun99}. As an initial model for the MHD simulations, we used
nonmagnetic, two-dimensional models whose computation was based completely on the
approaches presented in \cite{Gadun95}.

The upper and lower boundary conditions in the MHD simulations were taken to be
free; i.e., there was free inflow and outflow of material. The velocity components
were determined by the condition $\partial {\bf V}{/}\partial z = 0$, and the
average values of the intrinsic energy and density were fixed by the initial
uniform model \cite{Gadun95, Gadun96}. The profiles of the fluctuations of these
quantities at the upper (or lower) boundary coincide with the corresponding
variations for the layers located at lower (or greater) heights. In addition, the
density at the lower boundary was scaled so that the sum of the gas, radiative, and
magnetic pressures was constant at a fixed horizontal level.

The upper and lower boundary conditions for the magnetic field were specified by
its global character and took the form $B_x = 0$ and $\partial B_z{/}\partial z =
0$. The lateral boundary conditions were taken to be periodic. The initial
configuration of the magnetic field was bipolar, with the field intensity
decreasing with height. The average value of $B$ over the entire computational
region was 54~G. We chose this configuration to help ensure numerical stability of
the solution at the initial simulation time, since the velocity field and
thermodynamic quantities, on the one hand, and the magnetic-field characteristics,
on the other hand, are not self-consistent in the initial model.

The computational region was $3920 \times 1820$~km in size, with spatial steps of
35~km. The atmospheric layers covered about 700~km. It is obvious that we cannot
study the intrinsic structure of fine magnetic configurations in detail with such a
step size, but it is adequate for following trends in the evolution of magnetic
configurations, and reduces the computer time required for the calculation.

To study the evolution of magnetic elements, we considered a sequence of MHD models
from time 1.5 min at the beginning of the simulation (when the average magnetic
field was about 54~G) to time 120 min (when the average magnetic field was =
500~G). This sequence contains 94 two-dimensional models with time intervals of one
minute and 52 models with intervals of 0.5 min. The computational region contains
112 vertical columns (rays). We computed four Stokes profiles for each vertical
column for a particular model as for a plane-parallel atmosphere, after which we
performed a spatial average of the resulting profiles. The equations for the
transfer of polarized radiation in a magnetic field (often called the
Unno-Rachkovskii equations) form a system of four first-order differential
equations. We used a fifth-order Runge-Kutta-Felberg method to numerically solve
these transfer equations. The detailed computation algorithm, description of the
software, and behavior of the Stokes profiles as functions of the state of the
medium and atomic parameters are given in \cite{Landi76, Sheminova90, Sheminova91}.
The only substantial restriction in the computation of the Stokes profiles is the
assumption of local thermodynamic equilibrium.

To study temporal variations in the Stokes profiles, we chose the Fe I 15648~\AA\
infrared line, which possesses very promising diagnostic capability
\cite{Solanki92}. Along with its high magnetic sensitivity, the important
advantages of this line are its low temperature sensitivity and the fact that it is
formed in very deep layers of the photosphere (below $\log \tau_5 = -1$). As shown
in \cite{Sheminova99}, this line is especially suitable for LTE analyses of
theoretical MHD models, in which the temperature fluctuations in the upper layers
of the photosphere can be artificially overestimated due to the use of a gray
approximation for the radiative transfer.

\section{General characteristics of nonstationary \\ magnetogranulation
}

The results of the numerical simulations show that the character of the evolution
of thermal fluxes significantly affects the development of magnetic configurations.
The dissipation of granules and merging of intergranular gaps result in the
disappearance of small-scale magnetic fields or the formation of cluster-like
structures, whereas the fragmentation of granules in the presence of horizontal
photospheric fields leads to the formation of new compact magnetic configurations.

\begin{figure}
   \centering
   \includegraphics[width=13.cm]{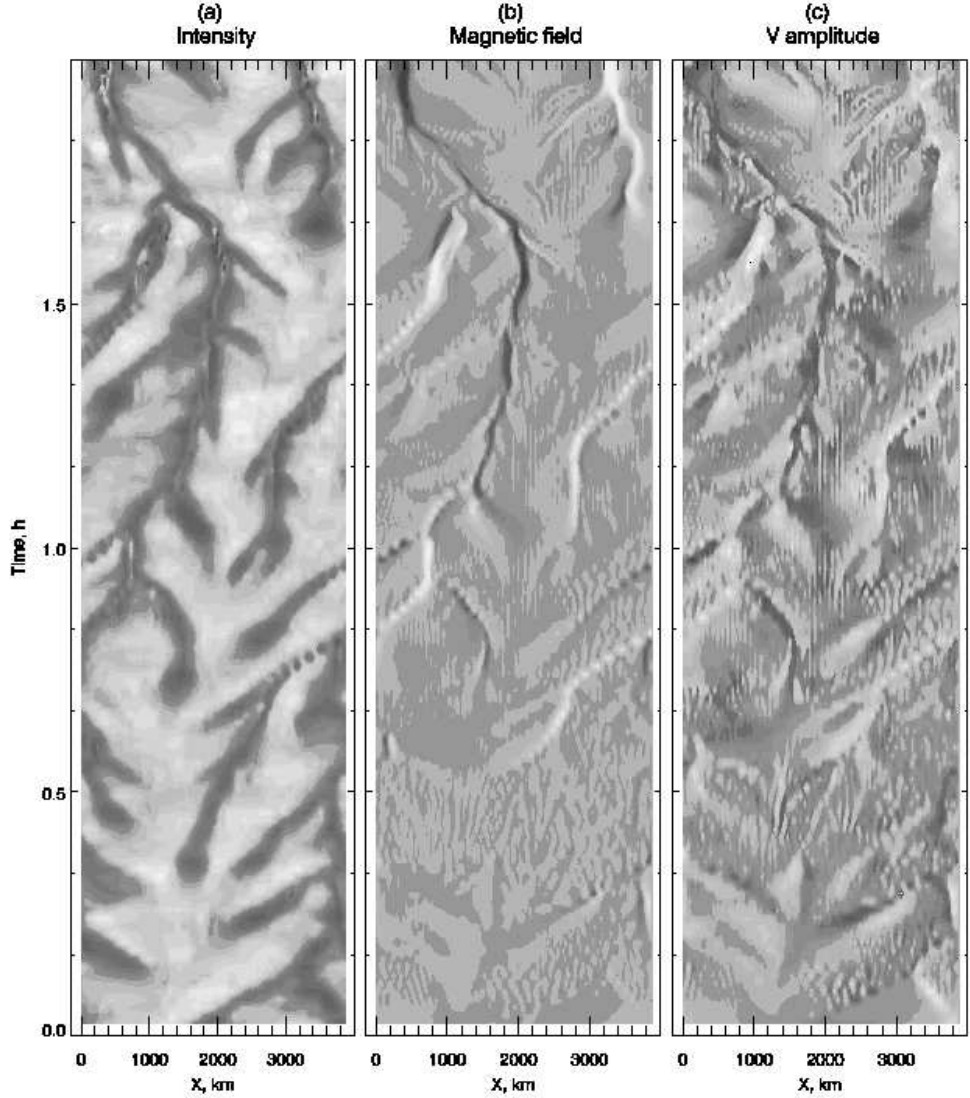}
   \caption[]{
Spatial and temporal evolution of (a) the intensity of monochromatic emission  $I_c/{<}I_c{>}$
at wavelength 500 nm, (b) the magnetic-field intensity $B$ at $\log \tau_R = 0$
level, and (c) the amplitude of blue wing of the Stokes V profile  of
the Fe~I 15648~\AA\ line  in the course of
two-hour, two-dimensional MHD modeling. The  $I_c/{<}I_c{>}$
variations  is 0.017-1.78, and is represented by a linear gradation from dark
to light shading. The  $B$  variations  ranges from -2867
to +2540 G. Negative and positive polarity of magnetic fields are shown by dark and
light shading, respectively.  The variations in the  V profile  amplitudes range from -0.14 to 0.16. Negative and positive
values are shown by dark and light shading, respectively.
 }
      \label{Fig1}
\end{figure}

Figure 1 shows that the entire simulation interval can be subdivided into three
stages \cite{Gadun00, Gadun99}: an initial period (up to 20 min), which is
primarily determined by the initial conditions; a transition period (from 20 to 35
min), when there is a mutual reconstruction of the magnetic field and thermal
convection; and a final period of self-consistent evolution of the nonstationary
convection and magnetic field. The first clear signatures of the presence of
compact magnetic tubes with strong fields (of the order of several kG) appear after
approximately 50 min of real solar time. Bright points form in the intergranular
gaps, and the tubes begin to glow. The positions of the bright points (Fig. la)
correspond to regions of strong magnetic field whose orientation is predominantly
vertical or close to vertical (Fig. 1b), and also to regions of increase in
temperature, decrease in pressure, and oscillating upward-downward flows in the
intergranular gaps.

Due to the effect of the hot walls, the glow of the tubes will begin earlier near
the disk edges than at their centers. This suggests the possibility of early
observational diagnostics of magnetic elements that have just begun to be formed
\cite{Brandt}. Note also that the compact structures with strong magnetic fields
that form are often located near regions with relatively weak fields of the
opposite direction. The character of the granulation also changes. The magnetic
field produces a stabilizing action: there are no strong horizontal shears in the
presence of the magnetic field, as distinct from the case of nonmagnetic
granulation. The size distribution of the granules in the magnetogranulation also
differs from the nonmagnetic case: the granules are smaller, and there are many
small-scale bright structures with sizes less than 150--300~km associated with
bright points of the magnetic tubes \cite{Gadun00}.

There are two scenarios for magnetic-tube formation in the numerical simulations
under consideration \cite{Gadun00}. The first is based on the concentration of a
magnetic field between cells and its intensification up to the equipartition level
due to a kinematic mechanism; the field can then be further amplified by the
development of superconvective instability. The second mechanism is based on the
formation of a tube during the decay of thermal convective flux. The magnetic-tube
behavior in this case also exhibits two stages. The first takes place when the
field intensity has not reached the equipartition level, so that the tube is in a
state of convective instability. The subsequent, stronger stage occurs when the
upper part of the tube is in radiative equilibrium and experiences an oscillatory
instability. The transition from one regime to the other occurs at $B =
1200$--1300~G \cite{Gadun00}.

The situation changes abruptly when convective collapse is "turned on." The level
of the observable surface of the magnetic tube becomes more strongly dependent on
$B$. The Wilson depression increases. The temperature at the corresponding level
also increases, since the observable surface shifts to deeper layers, and the
plasma configuration is actively heated by neighboring hotter thermal fluxes. The
gas pressure decreases due to the increasing magnetic pressure, and the density
also decreases. The tubes begin to glow, and the intensity of their emission is
above the average level. The tubes are also surrounded by dark regions, which
correspond to channels of intense downward flows of material near the tubes. The
brightness peaks of most wide tubes are split due to the effect of the large
horizontal scale, which results in less efficient lateral heating. As a result, the
total contrast of the tubes increases.

Our simulations revealed three basic mechanisms for the disintegration of the tubes
\cite{Gadun00}: dissipation of the magnetic field due to reconnection of the field
lines, disintegration due to interchange instability, and reversal of the
convective collapse. The first case---dissipation of the magnetic field---is the
most common scenario. It occurs when the thermal flux separating two tubes with
opposite field directions disappears, and the tubes merge. Interchange instability
takes place when the orientation of intense magnetic tubes deviates from the
vertical. The rising of strong horizontal magnetic field results in the development
of flute instability. In the plane case under consideration, this instability can
split and dissipate strong compact tubes. Inverse convective collapse
\cite{Gross98} is produced by the strong depletion of the upper part of the tube
when it is intensified. Such depletion reverses the convective collapse, so that
downward flows in the tube are replaced by upward ones, whose velocity can reach
supersonic values in the atmosphere \cite{Gross98}.

The evolution of magnetoconvection briefly outlined above is analyzed in more
detail in \cite{Gadun00}. Here, we will compare separate critical stages of tube
evolution with Stokes profiles synthesized for the same simulation times. With this
aim in view, we chose the most typical cases for the formation and decay of
magnetic tubes from the temporal sequence of MHD models.

\section{Examples of evolution of a kilogauss magnetic tube
}

\begin{figure}
   \centering
   \includegraphics[width=14.cm]{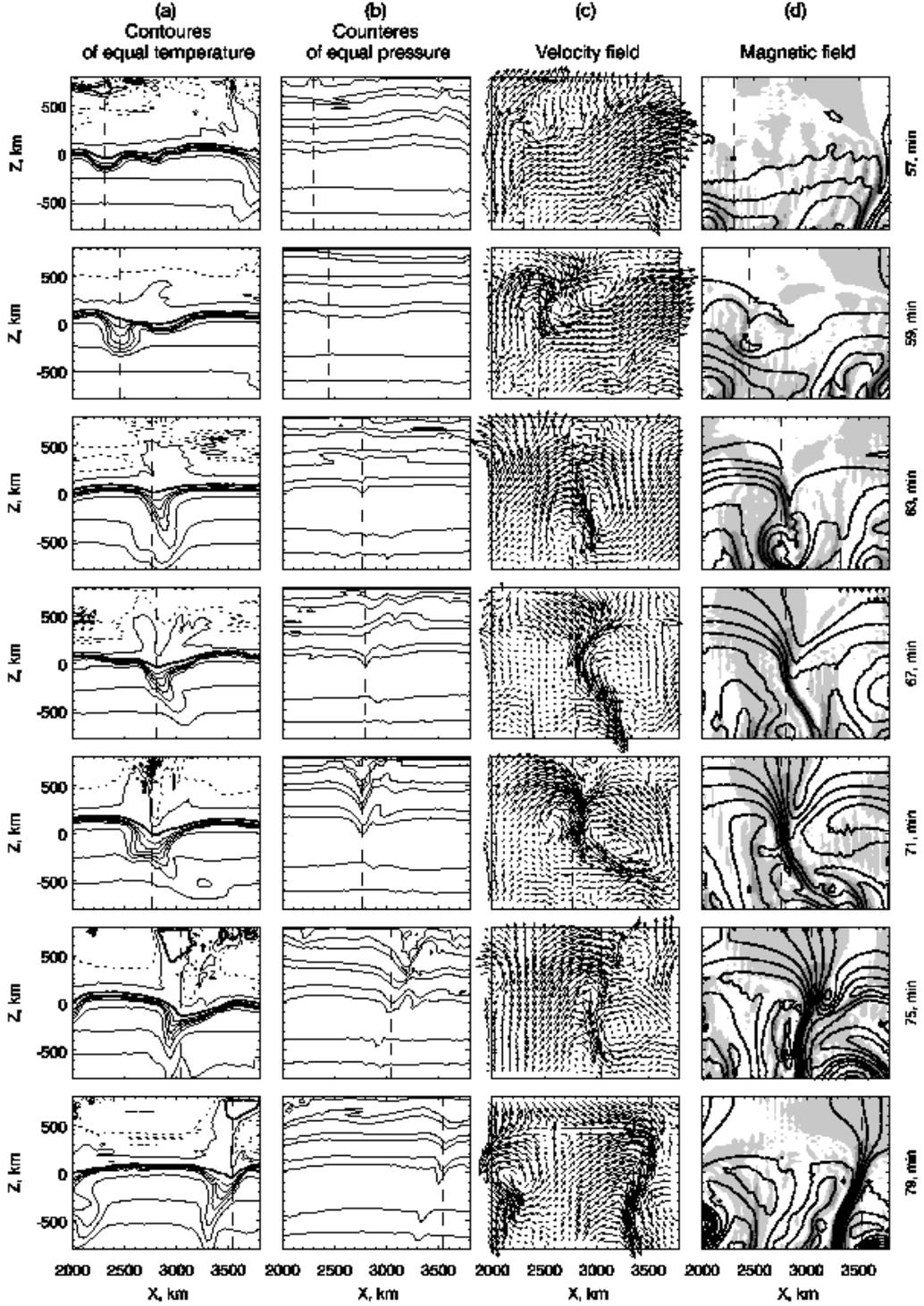}
   \caption[]{
Spatial and temporal evolution of magnetic-tube formation in the simulated region
from 2000--3800~km in the time interval 57--79 min: distribution of the (a)
temperature $T$ (the dotted curve corresponding to 4000 K and the thick curve to
6000~K show the  Fe I 15648~\AA\ line formation region), (b) gas pressure $P$, (c)
velocity field $V$, and (d) magnetic-field intensity $B$ (the density of the
shading is proportional to the values 10, 400, 800, and 1200~G; dark and light
regions correspond to positive and negative polarity, respectively). The vertical
dashed line denotes the center localization of the tube.
 }
      \label{Fig2}
\end{figure}

Figure 1c shows variations in the amplitudes of the blue wings of the synthesized
Stokes V profiles over two hours in the simulation region. These changes are very
similar to the space-time variations in the magnetic field (Fig.~1b). To illustrate
the formation and dissipation of the magnetic tubes, we shall consider in detail
only two separate regions of the MHD models, whose evolution is presented in
Figs.~2 and 3. The thick curve in the temperature distribution denotes the level
$\tau_R = 1$ (where $\tau_R $ is the Rosseland opacity), and the dotted curve
corresponds to the lowest temperature (4000 K). The vertical dashed line indicates
the central axis of the region in which the magnetic tubes are located. Figures 4
and 5 present Stokes profiles of the Fe I 15648~\AA\ spectral line with various
spatial resolutions with respect to the central tube axis --- 35, 105, 175, and 315
km --- synthesized for these regions. Note that a resolution of 35~km corresponds
to the horizontal step of the simulation, so that the corresponding profiles were
calculated at the tube center without spatial averaging; they are marked by the
solid curves in the figures. We determined the parameters commonly used in Stokes
diagnostics precisely for these profiles. These parameters are the amplitude of the
V profile, $a_V = (a_b + a_r){/}2$ (where $b$ and $r$ are the blue and red wings),
the magnetic-field intensity (in G) derived from the distance between the maximum
peaks of the V profile, $B_{br}= (\lambda\-r-\lambda_b){/}(2 * 4.67 \cdot 10^{-13}
\lambda_0^2 g_{\rm eff})$ (where $\lambda_0$ is the wavelength of the unshifted
line center in \AA), the shift of the V profile, $V_V = c(\lambda_z -
\lambda_0){/}\lambda_0 - 2.12\cdot10^{-6}c$ (where $\lambda_z$ is the wavelength of
the zero intersection of the profile), and the relative asymmetry of the amplitudes
and areas of the V profile, $\delta a = (a_b - a_r){/}(a_b + a_r)$ and $\delta  A
=(A_b-A_r){/}(A_b+A_r)$.

\begin{figure}[t]
   \centering
   \includegraphics[width=14.cm]{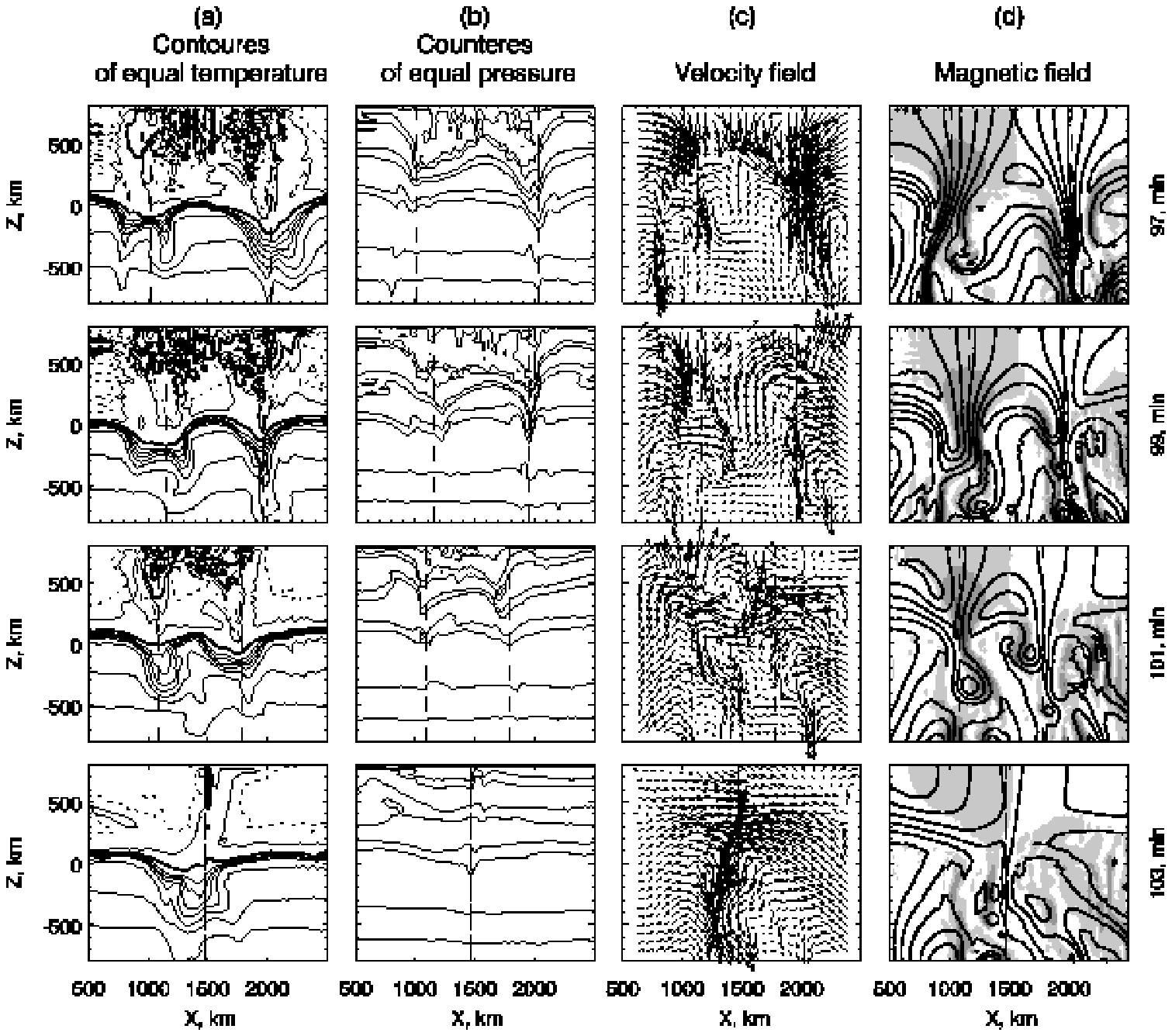}
   \caption[]{
Spatial and temporal evolution of the dissipation of two magnetic tubes in the
simulated region from 500--2500~km in the time interval 97--103~min. Format
is the same as in Fig. 2.
 }
      \label{Fig3}
\end{figure}

\subsection{Formation of a magnetic tube }

Let us compare the calculated Stokes profiles and their parameters (Fig. 4) with
the physical processes of magnetic-tube formation due to the surface mechanism
(Fig. 2).

A situation typical for the beginning of fragmentation of a large convective cell
with a diameter of about 3000~km occurs in the ($3920 \times 1820$~km)
computational region at time 57 min. There is an upward convective flow in its
central part ($x = 2100$~km). Two reverse flows, directed from the central regions,
can be seen at the periphery. A spot of high-density material appears in the upper
layers of the photosphere. The temperature at the  $\log \tau_R = 0$ level
decreases in regions of weak flow due to radiative cooling. There is a weak
horizontal magnetic field with varying direction in the cell. In the predicted
region of magnetic-tube birth (whose center is marked by the vertical dashed line
in Fig. 2), the calculated Stokes profiles are shifted toward the blue part of the
spectrum, $V_V = -0.8$~km/s. The minus sign denotes upward motion. The weak V
profile (drawn by the solid curve in Fig. 4) includes two components with opposite
polarities (the negative one is quite strong, and the positive one very weak).

\begin{figure}
   \centering
   \includegraphics[width=15.cm]{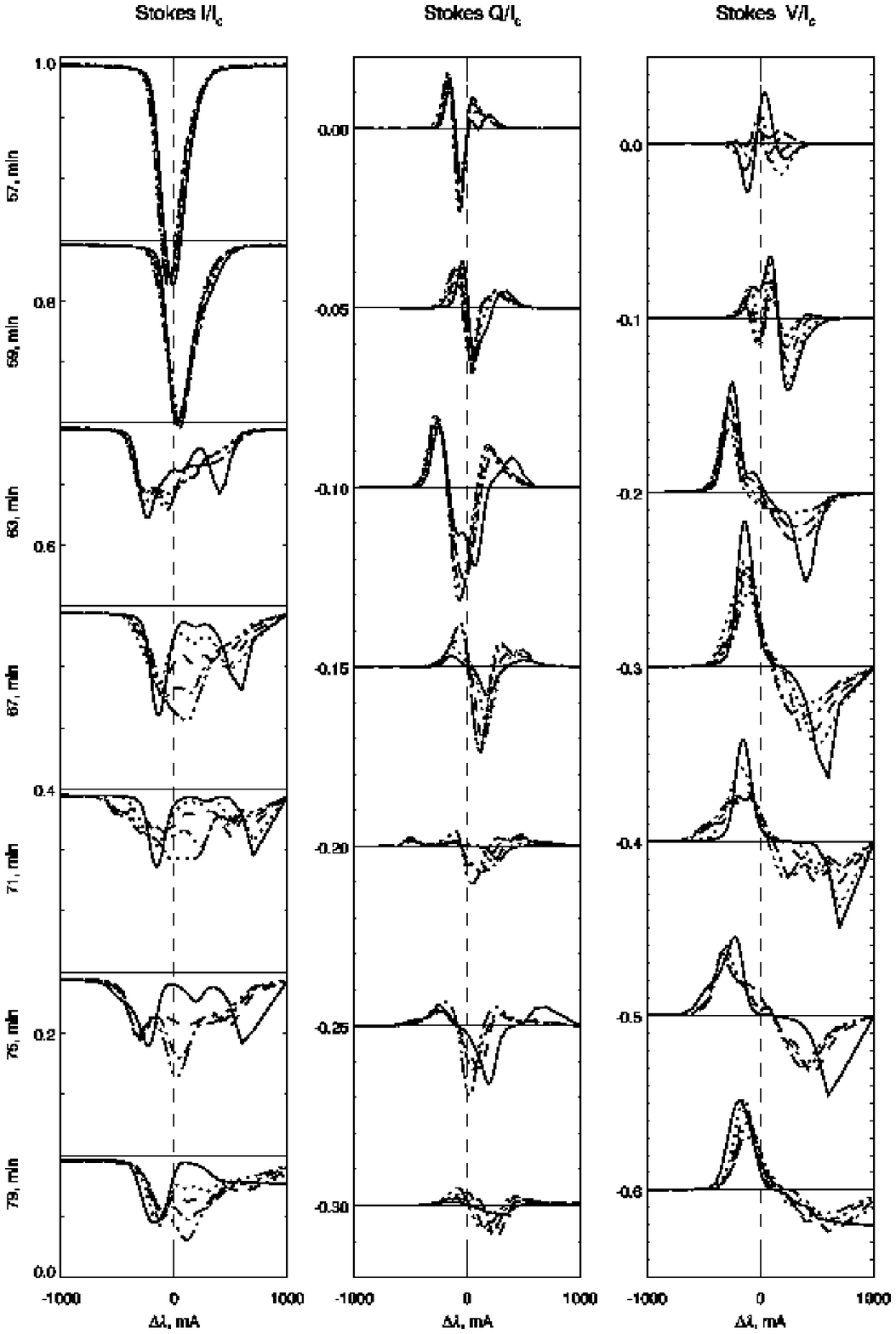}
   \caption[]{
Stokes profiles of the Fe I 15648~\AA\ line calculated at the center formation of
the kilogauss tube (denoted by vertical dashed line in Fig. 2) for various spatial
averaging scales: 35~km (solid curve), 105~km (dotted curve), 175~km (dashed
curve), and 315~km (dot-dash curve).
 }
      \label{Fig4}
\end{figure}

The situation at time 59 min is typical for the initial phase of magnetic-tube
birth. A downward jet of cold material forms during the fragmentation of the
thermal flux (a convective cell). This flow drags the photospheric magnetic field
with it, so that the horizontal field is transformed into a longitudinal field. The
I, Q, and V Stokes profiles are shifted toward the red part of the spectrum, and
$V_V = 3.0$~km/s. The changes in the shape of the V profile (where the positive
component is dominant) are the result of the increase in the longitudinal field
with positive polarity.

Further, the formation of the downward jet continues at times 63--67 min. The
stratification of the material increases. There is a deficit of the gas pressure,
and the positive magnetic field is sharply amplified. The accelerated vertical
flows carry cold material and the magnetic field downward. Hot walls are formed in
the upper part of the channel due to compression. The strong magnetic flux reaches
the bottom of the computational region. The amplitude of the V profile takes on its
maximum values. The complete splitting of the profiles indicates that the field
intensity in the tube reached kilogauss values ($B_{br} = 1078$~G). The redshift of
the profiles is 3.3~km/s.

The subsequent times 71, 75, and 79 min demonstrate the evolution of a fully formed
vertical magnetic tube with diameter 150~km at the $\log \tau_R  = 0$ level, which
is in a state of convective collapse. The kilogauss longitudinal magnetic field in
the channel enters deeper layers. The I Stokes profiles are completely split during
this interval ($B_{br} = 1246$, 1195, and 1231~G), the wing amplitudes have
decreased, and the redshifts are large --- $V_V = 4.9$, 2.2, and 3.4~km/s.

Thus, this fragment of magnetoconvection considered over an interval of 22 min
(from 57 to 79 min with respect to the beginning of the simulation) shows that the
surface mechanism for magnetic-tube formation reaches the stage of convective
collapse during its development. The downward flow of material amplifies the
magnetic flux in the channel to kilogauss values within 6--10 min. In the next 12
min, convective processes around the tube gather and push the magnetic field lines
toward the channel, leading to further concentration of the field in the magnetic
tube. These processes have an oscillatory character, due to the presence of both
global oscillations of the entire computational region and oscillations of the tube
by itself. In general, the characteristic spectral features of the evolution
revealed by the simulations confirm the results obtained in \cite{Gross98}.

\subsection{Decay of magnetic tubes }

Let us now consider a scenario for magnetic-field dissipation due to the forced
merging of two tubes with opposite field directions, which is most typical of the
simulations considered. The dynamics of these processes are shown in detail in Fig.
3, and the responses of the Stokes profiles calculated for each tube separately are
presented in Fig.~5.

\begin{figure}
   \centering
   \includegraphics[width=15.cm]{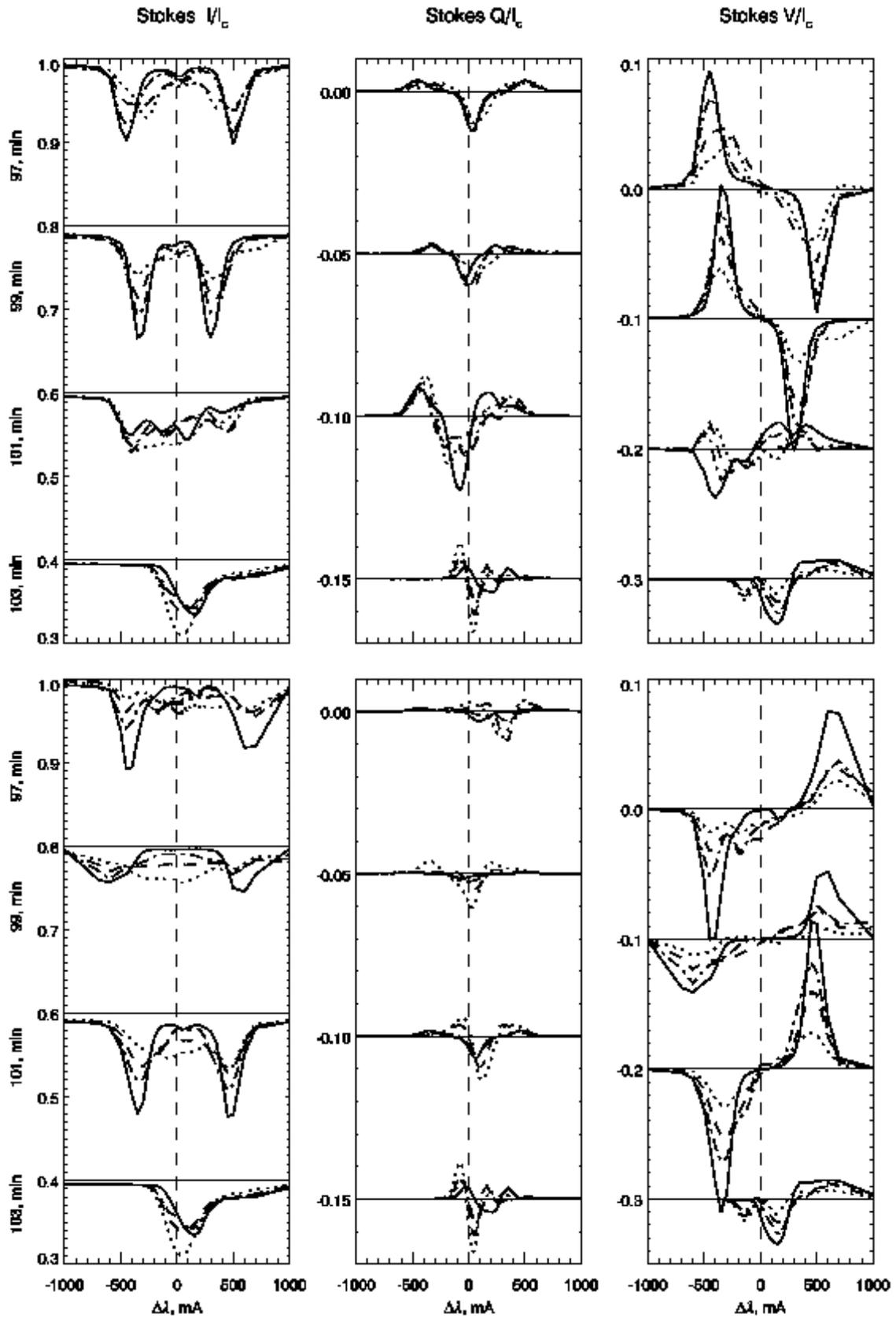}
   \caption[]{
Stokes profiles of the Fe I 15648~\AA\ line calculated at the center of the
kilogauss tubes (denoted by vertical dashed line in Fig. 3) in regions where there
is dissipation of these tubes. Top panels show Stokes I, Q, V profiles formed in a
tube with positive magnetic-field polarity and bottom panels show the profiles
formed in  another tube with negative polarity.
 All designations are the same as in Fig. 4.
 }
      \label{Fig5}
\end{figure}

We can see two kilogauss vertical tubes with different field directions at time 97
min, separated by a distance of about 1000~km. The disintegration begins in the
first, older, tube, which has a diameter of 350~km and positive polarity: the
concentration of the magnetic flux decreases due to the tube's separation and the
appearance of upward motions in deep layers inside the tube. The magnetic field has
already reached the bottom of the computational box; i.e., the kilogauss
longitudinal magnetic field has entered deep layers. The profiles are completely
split ($B_{br} = 1384$~G), have a regular shape, and are redshifted by 1.1~km/s.

The second tube, which has a smaller diameter (250~km) and negative polarity, is a
well-developed intense magnetic tube with some signatures of superconvective
instability: there are accelerated downward flows of material, a maximum
concentration of longitudinal magnetic field along the entire tube down to the base
of the model, a narrow region of downward motion with a gas-pressure deficit, a
rarefied region at the top, and a strong Wilson depression. At this time, the
distance between the peaks of the calculated V profile ($B_{br} = 1530$~G) and the
redshift ($V_V = 5.0$~km/s) are maximum.

At subsequent times 99--101 min, the thermal flux separating the tubes disappears,
and the two tubes with opposite magnetic-field directions begin to move toward each
other, resulting in reconnection of the field lines. Characteristic motions similar
to siphon flow can be seen between the tubes, whose features are described in
detail in \cite{Thomas91}. In the case under consideration, only magnetic-field
lines between the tubes (rather than in the entire tube volume) participate in the
siphon flows. This process decreases the magnetic field. In general, the dynamics
of the motion are very complex. The flows of material in the tubes have
considerably different velocity gradients and directions of motion. The upward flow
increases in the positive tube, while the downward flow dominates inside the
negative tube. Vortex motions appear between the tubes. All these processes result
in disintegration of the tube structure and, accordingly, a rapid decrease in the
magnetic-field concentration.

At time 101 min, the calculated V profile gives $V_V = -1.0$~km/s, $B_{br} = 700$~G
for the first tube and $V_V = 0.7$~km/s, $B_{br} = 1166$~G for the second tube. At
time 103 min, there is only a weak field with negative polarity in place of the two
kilogauss tubes. A new powerful flow of material begins to move downward along the
existing thermal channel with its pressure deficit. The profiles are appreciably
shifted toward the red part of the spectrum, and the right wing is very broad ($V_V
= 5.4$~km/s, $B_{br} = 656$~G).

Therefore, within a very short time interval of 6 min (from 97 to 103 min with
respect to the beginning of the simulation), dissipation of the thermal flux
(convective cell) separating two tubes with opposite field directions led to
merging of these compact magnetic elements and their disintegration.

\subsection{Evolution of magnetic tubes and variations \\ in the Stokes profiles }

To trace the dynamics of temporal variations in the calculated Stokes profiles in
the regions of formation and decay of the magnetic tubes, first and foremost, we
studied the influence of spatial averaging on the results of Stokes diagnostics.
This problem is considered in more detail in \cite{Sheminova99}. Figures 4 and 5
present the Stokes profiles calculated with various spatial averaging scales. A
comparison of these profiles clearly shows that their shape substantially depends
on the averaging region. The main reason for this is the complex structure of the
simulated magnetic tube. The presence of appreciable horizontal and vertical
gradients of the magnetic-tube parameters and of different directions for the
magnetic field in the averaging region result in an anomalous shape for the Stokes
profiles. The shapes are undistorted only for profiles calculated without
horizontal averaging, i.e., for a single column in the tube center. Thus, only such
profiles are able to reveal evolutionary spectral features. We used these profiles
to calculate the parameters of the Stokes diagnostics, which are presented in Fig.
6 as functions of time.

As can be seen from the dependences in the left-hand side of Fig. 6, the following
features are characteristic of the stage of magnetic-tube formation.

\begin{figure}[!th]
   \centering
   \includegraphics[width=11.5cm]{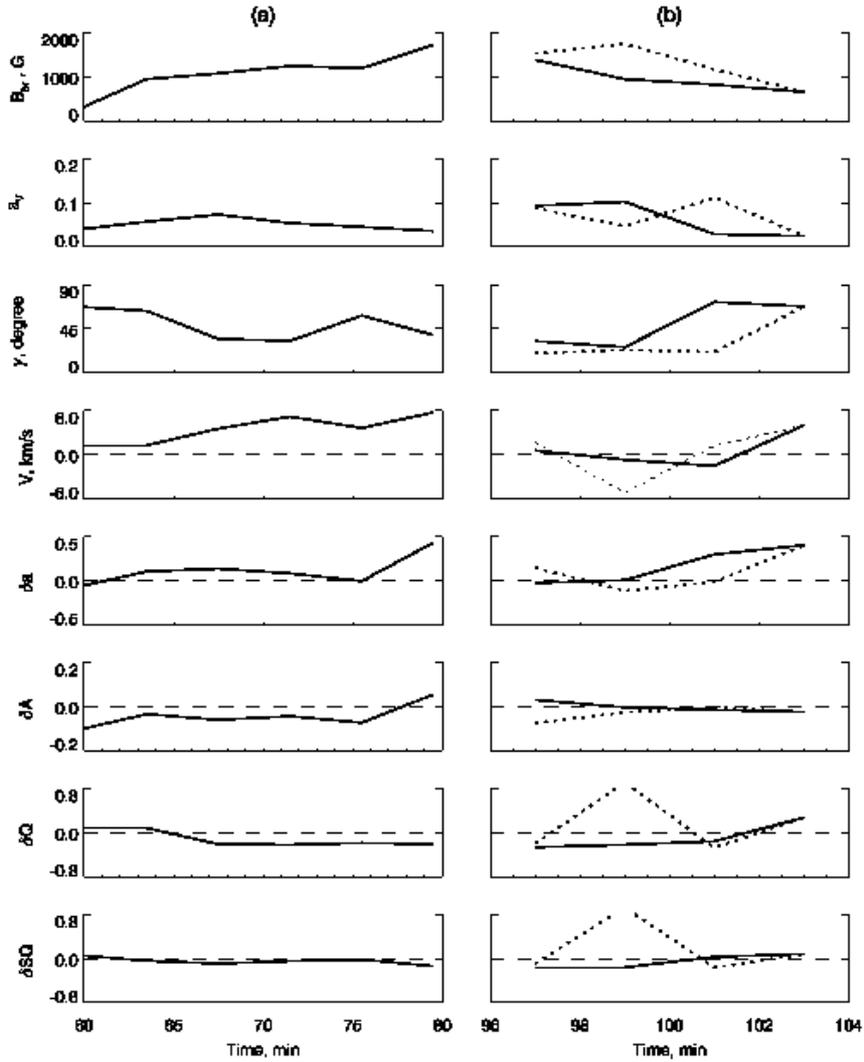}
   \caption[]{
Variations in the parameters of Stokes diagnostics during (a) formation of the
magnetic tube and (b) dissipation of two magnetic tubes. The solid and dotted lines
correspond to the tubes with positive and negative magnetic field polarity,
respectively. Here, $B_{br}$ is the magnetic-field intensity, $a_V$ is the
amplitude of the V profile, $V_V$ is the vertical velocity, and $\delta a$ and $
\delta A$ are the amplitude and area  asymmetries of the Stokes V.
 }
      \label{Fig6 }
\end{figure}

\begin{enumerate}
  \item  The distance between the $\sigma_{\pm}$, components of the
V profiles in units of the magnetic-field intensity $B_{br}$ reaches a maximum at
times 71--79 min, when the tube experiences convective collapse.
  \item  The amplitude of the V profiles, $a_V$, increases,
reaching its maximum just before the convective collapse, and then begins to
decrease.
  \item  Redshifts of the profiles $V_V$ predominate over the
entire formation period and lifetime of the magnetic tube. They are plotted in
velocity units in Fig. 6. The redshifts reach considerable values (up to 5~km/s)
during the convective collapse.
  \item  The amplitude asymmetry of the V profiles $\delta a$ is
positive, varies in the range 0.02--0.3, and is correlated with redshift.
  \item  For the most part, the area asymmetry $\delta A$ is negative
and considerably less than $\delta a$.
\end{enumerate}

The right-hand side of Fig. 6 presents the parameters of the Stokes diagnostics for
the tube decay. The positive tube (solid line) begins to disappear at time 97 min,
and the negative one (dotted line) at time 99 min. The resulting dependences can be
summarized as follows.

\begin{enumerate}
  \item The magnetic splitting of the profiles decreases.
  \item The   amplitude   of the   V profiles   initially
increases and then sharply decreases.
  \item  Blueshifts are dominant at the initial stage of disintegration
of the tube. They also can reach consider able values (up to 5~km/s), but are
observed for a sub stantially shorter time than are large redshifts.
    \item  The asymmetry of the V-profile amplitudes at the
beginning of the disintegration is close to zero or is negative, but later becomes
positive and increases.
      \item  The asymmetry of the V-profile areas slightly
decreases during the disintegration of the tube.
\end{enumerate}

The time dependences for the Stokes-profile parameters lead us to recommend
measurement of shifts of the zero intersection of the Stokes V profiles as a
practical diagnostic tool. Such shifts can be reliably determined from observations
with high spatial and temporal resolution.

\section{Conclusion
}

We have considered typical cases for the formation and decay of kilogauss magnetic
tubes in association with corresponding temporal variations in the Stokes profiles
of the Fe I 15648~\AA\ infrared line. The most characteristic signature of
convective collapse amplifying the magnetic field inside tubes to kilogauss values
is extremely large redshifts  of the Stokes V profiles and completely split I
profiles. On the contrary, the tube dissipation  process is characterized by
blueshifts of the Stokes profiles, with typical features of weakening fields.

{\bf Acknowledgements.} We are very grateful to S.K. Solanki for detailed
discussions of the results of this paper, valuable advice, and comments, and also
to S.R.O. Ploner for various assistance. This work was supported by the Swiss
National Scientific Foundation (grant no. 7UKPJ048440).


\end{document}